\documentclass[10pt,twocolumn,aps,pre,superscriptaddress]{revtex4-1}

\usepackage{amsmath}
\usepackage{graphicx}
\usepackage{color}
\usepackage{upgreek}
\usepackage{enumerate} 
\usepackage[linkcolor = blue, citecolor = blue, urlcolor = blue, colorlinks = true]{hyperref}
\usepackage[version=3]{mhchem}
\usepackage{commath,amssymb}
\usepackage{ushort}
\usepackage{multirow}
\usepackage[usenames,dvipsnames]{xcolor}
\usepackage{cleveref}
\usepackage{epstopdf}
\usepackage[exponent-product = \cdot]{siunitx}

\crefname{equation}{Eq.}{Eqs.}
\crefname{figure}{Fig.}{Figs.}
\crefrangeformat{equation}{Eqs.~#3(#1)#4--#5(#2)#6}

\newcommand{\siref}[1]{\textcolor{blue}{#1}}
\definecolor{myred}{RGB}{162, 20, 47}
\definecolor{myblue}{RGB}{0, 114, 189}
\definecolor{mygreen}{RGB}{119, 172, 48}
\definecolor{myyellow}{RGB}{237, 177, 32}

\begin{document}

\preprint{APS/123-QED}

\title{Autonomously Probing Viscoelasticity in Disordered Suspensions}

\author{Clara Abaurrea-Velasco}%
\email{c.abaurreavelasco@uu.nl}
\affiliation{Institute for Theoretical Physics, Center for Extreme Matter and Emergent Phenomena, Utrecht University, Princetonplein 5, 3584 CC Utrecht, The Netherlands}
\author{Celia Lozano}%
\affiliation{Fachbereich Physik, Universit{\"a}t Konstanz, Universit{\"a}tsstra{\ss}e 10, 78464  Konstanz, Germany}
\author{Clemens Bechinger}%
\affiliation{Fachbereich Physik, Universit{\"a}t Konstanz, Universit{\"a}tsstra{\ss}e 10, 78464  Konstanz, Germany}
\author{Joost de Graaf}%
\affiliation{Institute for Theoretical Physics, Center for Extreme Matter and Emergent Phenomena, Utrecht University, Princetonplein 5, 3584 CC Utrecht, The Netherlands}
   
\date{\today}

\begin{abstract}
Recent experiments show a strong rotational-diffusion enhancement for self-propelled microrheological probes in colloidal glasses. Here, we provide microscopic understanding using simulations with a frictional probe-medium coupling that converts active translation into rotation. Diffusive enhancement emerges from the medium’s disordered structure and peaks at a second-order transition in the number of contacts. Our results reproduce the salient features of the colloidal glass experiment and support an effective description that is applicable to a broader class of viscoelastic suspensions.
\end{abstract}

\maketitle
In active microrheology~\cite{waigh2005microrheology, squires2005simple, furst2017microrheology, gazuz2009active}, single particles are externally driven to probe the properties of the host medium. Inspired by this, self-driven colloidal ($\mu$m-sized) particles have recently been similarly employed~\cite{bechinger2016active, gomez2016dynamics, lozano2019active}. The autonomous nature of such active probes (APs)~\cite{bechinger2016active} gives rise to a truly novel microrheology, wherein the disordered environment not only induces a response dependent on local properties, but also guides the AP's exploration~\cite{lozano2019active}.

Experiments with APs in viscoelastic polymer suspensions~\cite{gomez2016dynamics} and quasi-2D glassy systems of bidisperse colloids~\cite{lozano2019active} reported orders-of-magnitude enhancement of the AP's rotational diffusion. This is remarkable considering the apparent difference in host medium. These observations are related to the underlying viscoelastic properties of the medium and can be heuristically understood in terms of a time-delayed mechanical response of the viscoelastic surrounding,~\textit{i.e.}, a memory effect~\cite{gomez2016dynamics, lozano2018run, narinder2018memory, lozano2019active}. Lozano~\textit{et al.}~\cite{lozano2019active} indeed showed that rotational-diffusion enhancement (RDE) peaked around the glass transition, where the relaxation time of the suspension is maximum. Key to these experiments is that the probe does not significantly perturb the environment, which sets these apart from other scenarios involving active particles and the glass transition~\cite{berthier2013non, ni2013pushing, berthier2014nonequilibrium, nandi2018random}. Recent simulations~\cite{qi2020enhanced} gave insight into the role of polymer-AP interactions on the enhancement observed by Gomez-Solano~\textit{et al.}~\cite{gomez2016dynamics}. However, there is presently no microscopic understanding of what underlies the RDE in the glassy system.

In this Letter, we present a simple simulation model, wherein active translation is converted to rotation through a frictional contact coupling between the AP and its surrounding. Our description has no intrinsic memory effect, commonplace in other frictional models,~\textit{i.e.}, our model is Markovian on the single-particle level. Nonetheless, a time-delayed mechanical response emerges from the AP's interaction with the viscoelastic environment, which is stronger for softer potentials. Together with the local disorder of the colloidal suspension, this is sufficient to reproduce the spiked and asymmetric RDE of Ref.~\cite{lozano2019active}.

\begin{figure}[!htb]
\centering
\includegraphics[width=\columnwidth]{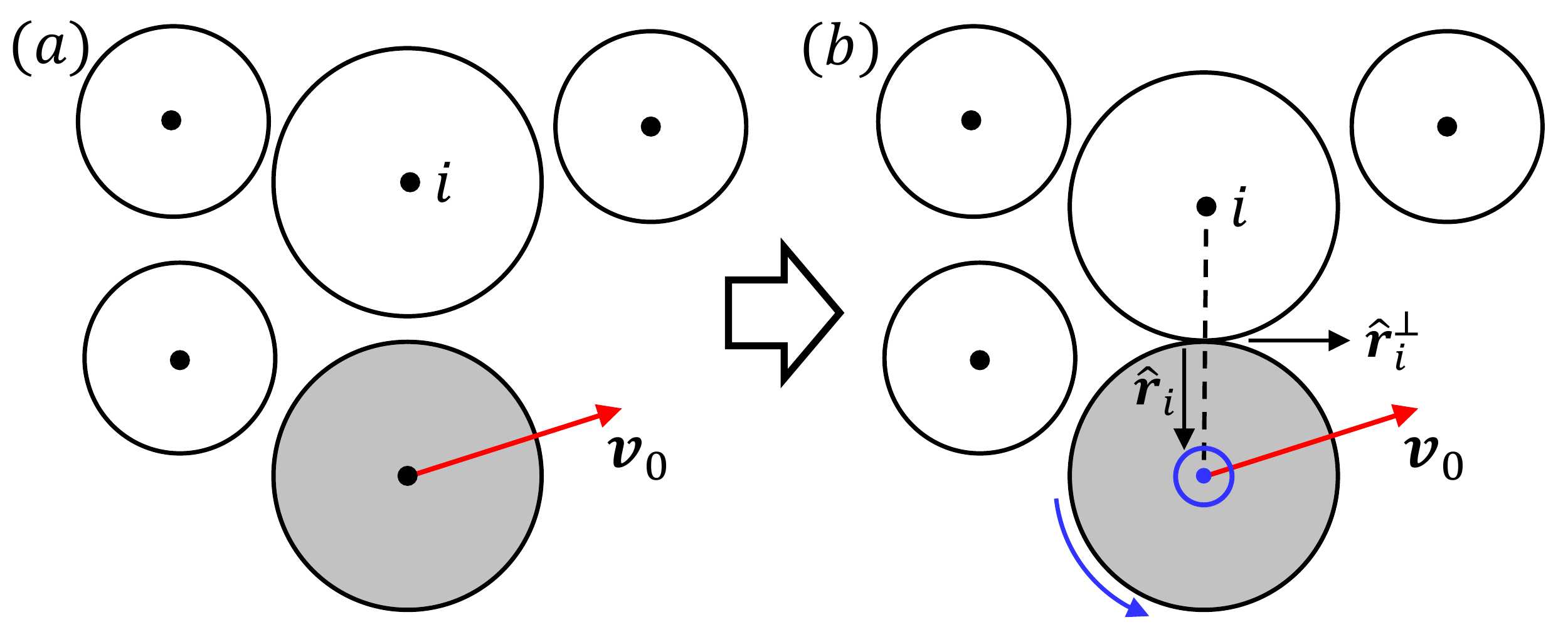}
\caption{\label{fig:friction_sketch} Active probe (AP; gray) in contact (dashed lines) with passive particles (white). Switching from (a) to (b) frictional coupling occurs, as a contact is established with the $i$-th neighbor. The red arrow indicates the AP's self-propulsion velocity ($\boldsymbol{v}_{0}$) and the two black arrows the normal $\hat{\boldsymbol{r}}_{i}$ and tangential $\hat{\boldsymbol{r}}_{i}^{\perp}$ contact direction. Contact induces an angular velocity ($\boldsymbol{\omega}_{i}$;  \textcolor{blue}{$\odot$}) pointing out of the plane, in turn leading to reorientation (blue arrow).}
\end{figure}

\begin{figure*}[!htb]
\centering
\includegraphics[width=2\columnwidth]{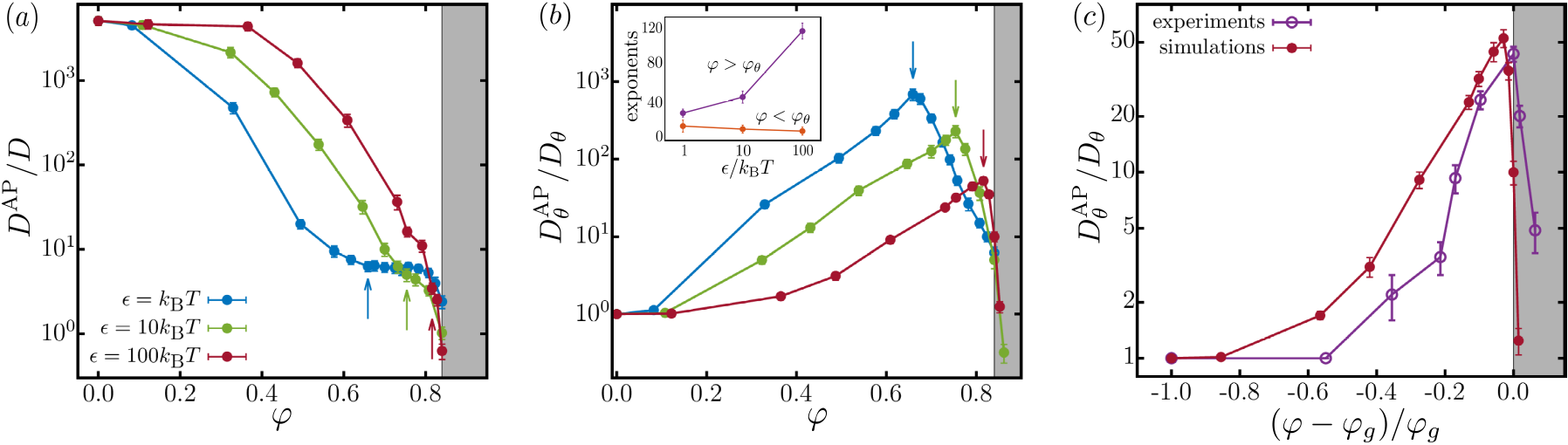}
\caption{\label{fig:dt-dr}AP translational and rotational diffusion in the glassy background. a)~Normalized AP translational diffusion coefficient $D^{\mathrm{AP}}$ versus area fraction $\varphi$; $D$ is the passive bulk coefficient. b)~Normalized AP rotational diffusion coefficient $D_{\theta}^{\mathrm{AP}}$ \ versus $\varphi$; $D_{\theta}$ is the rotational diffusion coefficient in bulk. The inset shows the fitted exponents to the left and to the right of $\varphi_{\theta}$,~\textit{i.e.}, the peak area fraction. The arrows in (a,b) show $\varphi_{\theta}$ and the colors are as indicated in the legend to (a). c) Comparison of $D^{AP}_{\theta}/D_{\theta}$ between experiment~\cite{lozano2019active} and $\epsilon=100k_{\mathrm{B}}T$ simulations as a function of the distance to the glass transition $\varphi_{g}$. The gray area marks the region beyond the glass transition in all plots. The error bars show the standard error of the mean.}
\end{figure*}

We closely matched the experiment of Ref.~\cite{lozano2019active} in our Brownian dynamics simulations with regular Gaussian noise,~\textit{i.e.}, without memory. We prepared quasi-2D systems of bidisperse passive particles with 1:1 stoichiometric ratio with respective diameters $\sigma$ and $0.698\sigma$; size ratio 1.4. These repelled each other \textit{via} a Weeks-Chandler-Andersen (WCA) interaction with strength $\epsilon$. We examined the effect of WCA softness by considering $\epsilon = 1$, $10$, and $100k_{\mathrm{B}}T$ with $k_{\mathrm{B}}$ the Boltzmann constant and $T$ the temperature. Our units of length, time, and energy were $\sigma$, $\tau=\sigma^{2} D^{-1}$, and $k_{\mathrm{B}}T$, respectively, with $D$ the bulk translational (and $D_{\theta}$ the 2D rotational) diffusion of the largest species. We simulated between 274 and 3014 particles in a periodic square simulation box with edge length $L=40\sigma$, corresponding to area fractions $\varphi \in [0.1,0.915]$. Here, $\varphi = (N/2) \pi \left( (\tilde{\sigma}_{\epsilon}/2)^{2} + (0.698\tilde{\sigma}_{\epsilon}/2)^{2} \right)/L^{2}$ with $\tilde{\sigma}_{\epsilon}$ the effective hard-core diameter of the soft particles, obtained from the radial distribution function, see Fig.~\siref{S2}~\cite{supplement}. Standard self-intermediate scattering function (SISF) analysis, see Fig.~\siref{S1}~\cite{supplement}, revealed that the systems underwent a glass transition at $\varphi_{g} \approx 0.84$ independent of $\epsilon$, in line with the hard-core normalization of $\varphi$. The particle displacements also showed a strong non-Gaussianity at $\varphi_{g}$, see Fig.~\siref{S9}~\cite{supplement}. Full details of the preparation procedure are provided in Section~\siref{S2}~\cite{supplement}.

After equilibration, we randomly selected a single large passive particle and turned this into an AP, sampling for times up to $t = 3 \cdot 10^{3}\tau$. Our AP was modeled as an active Brownian particle with a self-propulsion velocity given by $\boldsymbol{v}_{0}$, see Section~\siref{S5}~\cite{supplement}. Here, the speed $|\boldsymbol{v}_0| = \mathrm{Pe} \, \sigma D_{\theta}$ with $\mathrm{Pe}$ the P{\'e}clet number; we maintained $\mathrm{Pe}=120$ throughout to best approach the conditions of the experiment~\cite{lozano2019active}. We imposed that the AP interacted with its passive neighbors through a pair-wise frictional coupling, which transformed self-propulsion into reorientation:
\begin{align}
\label{eq:convert} \boldsymbol{\omega}_{i} &= 2\frac{\beta_{R}}{\sigma} \left( \hat{\boldsymbol{r}}_{i} \times (\hat{\boldsymbol{r}}_{i}^{\perp} \cdot \boldsymbol{v}_{0}) \hat{\boldsymbol{r}}_{i}^{\perp} \right) ,
\end{align}
with $\boldsymbol{\omega}_{i}$ the angular-velocity contribution due to contact with the $i$-th neighbor, $\beta_{R}$ the coupling coefficient, and $\hat{\boldsymbol{r}}_{i}$ and $\hat{\boldsymbol{r}}_{i}^{\perp}$ the normal and tangent contact unit vectors, respectively, as defined in Fig.~\ref{fig:friction_sketch}. We maintained $\beta_{R} = 0.5$ throughout for convenience. Angular velocity was generated when the distance between the AP and a Brownian neighbor was less than the WCA cut-off range,~\textit{i.e.}, $2^{1/6}$ times the mean diameter of the pair, see Fig.~\ref{fig:friction_sketch}.

In formulating the model, we took inspiration from recent work on the effects of friction in jammed systems~\cite{wyart2014discontinuous, seto2013discontinuous, mari2014shear, guy2015towards, hsu2018roughness, guy2020testing, singh2020shear}. However, unlike the standard granular friction model~\cite{cundall1979discrete, luding2005anisotropy}, we did not incorporate memory effects into the generation of $\boldsymbol{\omega}_{i}$,~\textit{i.e.}, our single-particle dynamics is Markovian. Our neighbor criterion encodes that reorientation occurs against a sufficiently rigid (solid) background, when the separation is small. The data provided in the main text assumes two additional contacts per passive particle for coupling to occur --- we took inspiration from the frictional isostaticity criterion in granular systems --- but the modeling is qualitatively robust to such a change, see Section~\siref{S5}~\cite{supplement}.

We obtained the AP's translational, $D^{\mathrm{AP}}$, and rotational, $D_{\theta}^{\mathrm{AP}}$, diffusion coefficients from the respective long-time mean-squared displacements $\langle \Delta r^{2}(t) \rangle = 4 D^{\mathrm{AP}} t $ and $\langle \Delta \theta^{2} (t)\rangle = 2 D_{\theta}^{\mathrm{AP}} t $, see Figs.~\siref{S4} and~\siref{S5}~\cite{supplement}. Figure~\ref{fig:dt-dr}a shows that $D^{\mathrm{AP}}$ decreased upon approaching the glass transition, due to caging effects  in the passive medium,~\textit{i.e.}, the active particle weakly perturbs the passive surrounding and cannot effectively push neighbors out of the way. Our neighbor criterion led to a strong correlation between translation and rotation of the AP at sufficiently high $\varphi$, see Fig.~\siref{S11}~\cite{supplement}, similar to the experiment~\cite{lozano2019active}. For $\epsilon = 1k_{\mathrm{B}}T$, the coupling caused $D^{\mathrm{AP}}$ to plateau below the area fraction, $\varphi_{\theta}$, for which the maximal RDE was found (vertical arrows). That is, active forward motion is almost entirely converted into rotation. A similar plateau was observed in experiments~\cite{lozano2019active}; in the simulations the plateau shrunk with increasing $\epsilon$.

We observed a peak $D_{\theta}^{\mathrm{AP}}$ as a function of $\varphi$, see Fig.~\ref{fig:dt-dr}b. The RDE proved to be systematically larger for softer potentials, with the peak moving to lower $\varphi$. The greater enhancement was the result of a larger viscoelastic response in the low-$\epsilon$ systems, see Fig.~\siref{S10}~\cite{supplement}. The shift in the peak is correlated with a systematic increase in the number of neighbor contacts with decreasing $\epsilon$, as will be discussed shortly. We also found an asymmetry of the exponential decay of the RDE away from $\varphi_{\theta}$, see the inset to Fig.~\ref{fig:dt-dr}b, reminiscent of the experimental trend~\cite{lozano2019active}; the fit procedure by which the exponents were obtained is detailed in Section~\siref{S7}~\cite{supplement}. For all $\epsilon$, the exponent by which the RDE increased for $\varphi < \varphi_{\theta}$, was the same within the error (standard error of the mean always). This makes sense, the coupling mechanisms is the same for all $\epsilon$ and the AP is in a regime where it can readily push passive particles out of the way and sample neighborhoods, as we will return to in the discussion. The coefficient of RDE decay ($\varphi > \varphi_{\theta}$), however, increased non-linearly with $\epsilon$. This is because RDE reduction is due to the AP staying longer in the same neighborhood, as the system becomes increasingly dense and stiff. Lastly, our simple choice for the coupling~\eqref{eq:convert} led to a linear velocity dependence of the maximal RDE. The quadratic experimental dependence of the RDE on the AP velocity may be reproduced through a speed-dependent coupling: $\beta \propto \vert \boldsymbol{v}_{0} \vert$, which is representative of a load-based coupling~\cite{followup}.

We tested our model's ability to capture the experimental RDE by comparing to Ref.~\cite{lozano2019active}, see Fig.~\ref{fig:dt-dr}c. Here, we chose $\epsilon=100k_{\mathrm{B}}T$ as this data closest matched the experiment. We accounted for the difference in the location of the glass transition between experiment ($\varphi_{g} \approx 0.776$) and simulation ($\varphi_{g} \approx 0.84$) through a normalized area fraction $(\varphi-\varphi_{g})/\varphi_{g}$. The agreement is excellent in view of limited parameter tuning. We have kept our modeling minimal with an outlook on a general viscoelastic fluids, possible refinements are listed in the discussion.

\begin{figure}[!htb]
\centering
\includegraphics[width=0.7\columnwidth]{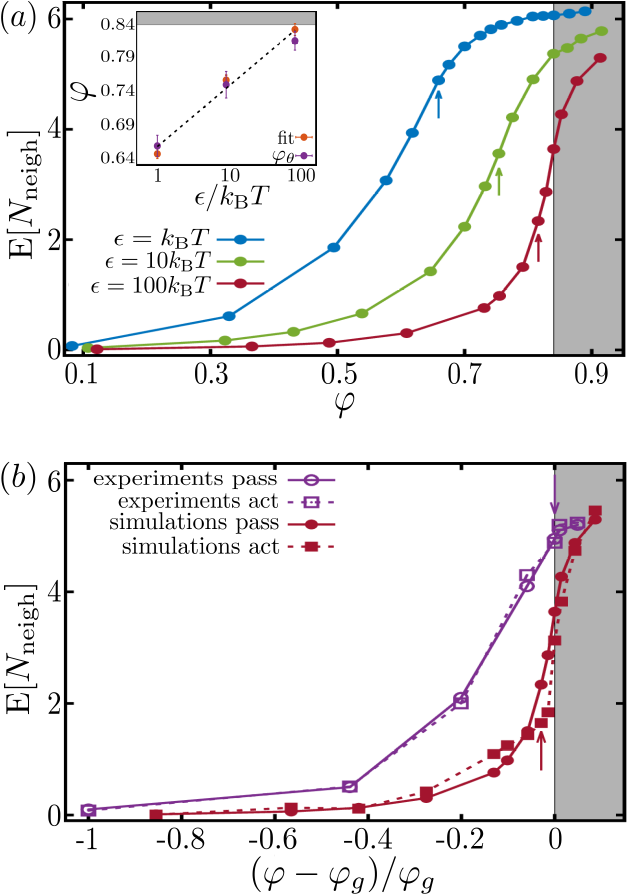}
\caption{\label{fig:expec-ang}Quantification of the local structure of the environment for passive systems and ones with an AP. a) Expectation value of the number of neighbors $\mathrm{E}[N_\mathrm{neigh}]$ surrounding a passive particle of size $\sigma$ versus $\varphi$. The inset shows the inflection point from the expectation value curves and $\varphi_{\theta}$ versus the softness of the WCA potential $\epsilon$ in terms of thermal energy $k_{\mathrm{B}}T$. b) Comparison of $\mathrm{E}[N_\mathrm{neigh}]$ for passive particles of size $\sigma$ and the AP between experiment~\cite{lozano2019active} and $\epsilon=100k_{\mathrm{B}}T$ simulations as a function of the distance to the glass transition $\varphi_{g}$. The gray area marks the region beyond the glass transition in both plots and the vertical arrows show $\varphi_{\theta}$.}
\end{figure}

In the experiment, care was taken to probe the linear-response regime,~\textit{i.e.}, the AP only weakly perturbed the local structure; a condition that we approached in our simulations. We therefore anticipated that a structural transition in the passive system determines $\varphi_{\theta}$. We tested this hypothesis by computing the probability density function (PDF) of the number of neighbors (soft contacts), $N_{\mathrm{neigh}}$, surrounding passive particles of size $\sigma$ (same as the AP), see Figs.~\siref{S6}-\siref{S8}~\cite{supplement}. The inset to Fig.~\ref{fig:expec-ang}a shows that for the 3-contact criterion the location of the inflection point of the associated expectation value $\mathrm{E}[N_{\mathrm{neigh}}]$ (orange) coincided with $\varphi_{\theta}$ (purple) for all $\epsilon$; a clear structural signal for $\varphi_{\theta}$ and is robust to our contact criterion, see Section~\siref{S5}~\cite{supplement}. The inflection-point fit procedure is outlined in Section~\siref{S7}~\cite{supplement}. Figure~\ref{fig:expec-ang}b compares the local structure around active and passive particles between experiment and simulation. The main difference is the steepness of the curves, yet, within the error, the experimental RDE peak also lay at the inflection point. The importance of an \textit{inflection} point prompts us to suggest that the associated structural transition is second order.

\begin{figure}[!htb]
\centering
\includegraphics[width=0.9\columnwidth]{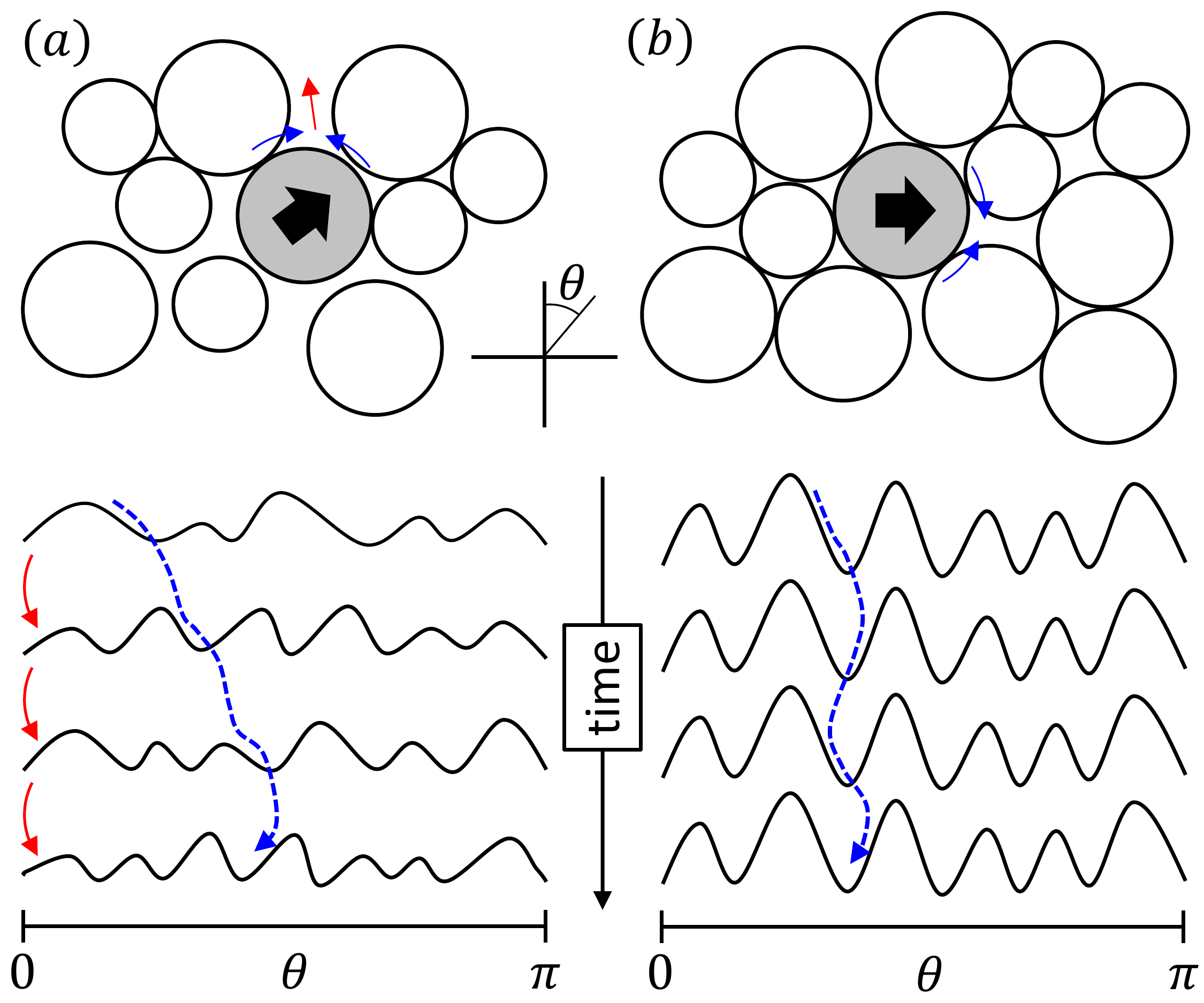}
\caption{\label{fig:rotation-sketch}Sketch of the RDE process. The top shows the AP (gray; oriented along the black arrow) probing its local environment for weak points. Blue arrows indicate active reorientation toward a possible escape (red arrow). The bottom rows show the emergent reorientation potential surrounding the AP, with angle $\theta$ measured with respect to the $y$-axis; peaks correspond to particles and valleys to the gaps between. (a) The AP hops from one local environment to another (red arrows) reorienting in each neighborhood ($\varphi \approx \varphi_{\theta}$), which results in a cumulative change in angle (blue dashed arrow). (b) The local environment is too crowded for the AP to escape ($\varphi \gtrsim \varphi_{g}$) and it only reorients weakly.}
\end{figure}

We are now in a position to understand how RDE occurs when there is a frictional coupling; this concept is illustrated in Fig.~\ref{fig:rotation-sketch}. In a dilute environment ($\varphi \ll \varphi_{\theta}$), the AP intermittently encounters passive neighbors and these are predominantly pushed out of the way. That is, there is no or very limited active reorientation, a feature captured in our model \textit{via} our neighbor criterion. When the local environment becomes more crowded and stiffer, the AP pushes itself off against its neighbors and can actively reorient. This can be pictured as angular motion in an effective `orientational potential'. This potential can be modified by the AP as it probes the local environment for weak points, by which it can escape, or through thermal fluctuations that rearrange the neighbors.

For deformable neighborhoods ($\varphi < \varphi_{g}$), the AP can easily overcome the barriers and hop to another neighborhood reorienting in each new orientational potential, see Fig.~\ref{fig:rotation-sketch}a. This process repeats itself and leads to continuous reorientation that averages out on sufficiently long time scales to enhanced rotational diffusion. The AP does not couple sufficiently to the environment to induce a truly long-time persistence, as has been found in other experimental systems~\cite{narinder2018memory}. However, for larger APs than considered here at sufficiently high velocities, we expect our model to recover such persistent rotation.

Orientational memory emerges by the way the AP moves in the orientational potential,~\textit{i.e.}, toward a local weak spot. This expresses itself as a decaying reorientational persistence with clear signs of correlation, see Fig.~\siref{S12}~\cite{supplement}. Conversely, in a very dense environment ($\varphi > \varphi_{g}$), the AP is trapped by surrounding neighbors, see Fig.~S7 that shows a homogeneous caging. In this case, the AP will orient toward a local weak spot and become stuck, see Fig.~\ref{fig:rotation-sketch}b. This can be understood as an orientational potential with high barriers. The limited per-cage active reorientation is smeared out over the large times between hops, leading to an insignificant RDE.

The RDE peak ($\varphi \approx \varphi_{\theta}$) is located where the amount of reorientation per environment and hopping rate complement each other. Figure~\siref{S11} shows a maximal correlation between (active) displacement and reorientation at $\varphi = \varphi_{\theta}$~\cite{supplement}. This is related to the value of $\varphi$ for which the average angular distribution of AP-neighbor contacts rapidly becomes homogeneous, see Fig.~\siref{S7}, which is commensurate with the rapid rise in $\mathrm{E}[N_{\mathrm{neigh}}]$. That is, the more homogeneous the environment, the weaker the enhancement, as pair-wise contributions cancel each other out and the gaps between particles become increasingly stable points of the orientational potential. This also leads to a change in trend for the emergent viscoelasticity as a function of $\varphi$, see Fig.~\siref{S10}~\cite{supplement}.

Our simple model captures the salient features of the experiment and provides microscopic insights into the origin of the RDE. However, the differences also prove informative. Starting with the passive system, we observe that the experimental glass transition was found at $\varphi_{g} \approx 0.776$ and displayed a sharp change of the SISF from fluid-like to arrested~\cite{lozano2019active}. Our otherwise matched simulations revealed $\varphi_{g} \approx 0.84$ and a broader change in the SISF. Increasing $\epsilon$ would sharpen the transition, but not affect the location of $\varphi_{g}$. It would also eliminate the RDE, extrapolating the trend in our data, which is the crucial experimental feature. That is, the RDE requires viscoelasticity of the environment, which systematically decreases with increasing $\epsilon$ at $\varphi_{g}$, see Fig.~\siref{S10}~\cite{supplement}.

Another important difference between simulation and experiment is that the RDE peak was found before the glass transition in the former. This is because particles come into contact at the edge of the soft potential, leading to early enhancement. However, RDE-reducing stiffness and homogeneity of the environment set in well before the dynamics becomes fully arrested, see Figs.~\siref{S10} and~\siref{S7}, respectively, from the SI~\cite{supplement}.

We suspect that the apparent dichotomy may be resolved by introducing contact friction between the passive colloids near the glass transition,~\textit{i.e.}, a more complete friction-based modeling than pursued here. Sliding friction between passive particles would lead to an earlier onset of arrest and the transition in neighbor contacts could force the glass transition to coincide with it. The reduced $\varphi$ for full dynamic arrest, would also give the experimental (almost) hard-sphere particles sufficient wiggle room to move and exhibit RDE~\cite{lozano2019active}, as at the simulated glass transition there is virtually no probe displacement, see Fig.~\ref{fig:dt-dr}a. Local contacts in the dense fluid and potential sliding-friction that these induce, could also explain the broadening of $\mathrm{E}[N_{\mathrm{neigh}}]$ and the presence of the plateau in $D^{\mathrm{AP}}$ with respect to the simulation, see Figs.~\ref{fig:expec-ang}b and~\ref{fig:dt-dr}a, respectively.

The above should be considered in the light of recent experiments showing the relevance of contacts to the aging of colloidal glasses~\cite{bonacci2020contact} and the known effect of friction on the location of the athermal shear thickening and jamming transition~\cite{vaagberg2014dissipation, comtet2017pairwise}. Hydrodynamic interactions have also been considered in this context and might lead to similar effects~\cite{ladd1995temporal, wang2020hydrodynamic}. However, the experiment~\cite{lozano2019active} did not reveal a significant change in the passive orientational diffusion even at $\varphi_{g}$, which suggests that lubrication-based dampening played a limited role. If friction is indeed present in experimental colloidal glass formers~\cite{lozano2019active, li2020anatomy, bonacci2020contact}, this would limit their fundamental connection to idealized hard-sphere systems, but also offers new richness. This will be the subject of future study.

Lastly, we consider the experiments on APs that exhibited RDE in a viscoelastic polymer suspension~\cite{gomez2016dynamics}. Based on our modeling, we expect that for very soft particles (polymer blobs) RDE can be observed at low dilution,~\textit{i.e.}, our simple description is capable of capturing RDE in such a viscoelastic fluid. A recent multi-particle collision dynamics study using a self-propelled particle (hydrodynamic squirmer) and model polymers~\cite{qi2020enhanced} indicated three key ingredients: (i) activity induced polymer desorption, (ii) asymmetric encounters with polymers, and (iii) heterogeneity of the polymer suspension. In our model, (iii) is covered by the bidisperse particle size, though we expect that disorder is generally sufficient; (i) and (ii) lead to a net reorientation. The strength of our approach lies in its simplicity, consequent computational efficiency, and generality.

In conclusion, we have provided a general microscopic understanding of the experimentally observed rotational-diffusion enhancement of active probes~\cite{gomez2016dynamics, lozano2019active}. In our model, activity-based, frictional coupling of linear and angular motion conjoins with the underlying structural characteristic of the disordered colloidal fluid that exhibits a viscoelastic response, to bring about the effect. Maximal enhancement occurs when there is a second-order transition in average `soft' neighbor contacts. This insight should prove critical for future experiments involving autonomous microrheological probes in complex environments, such as active (biological) fluids.

JdG gratefully recognizes the Dutch Research Council (NWO) for funding this research through Start-Up Grant 740.018.013. CB acknowledges financial support by the ERC Advanced Grant ASCIR (Grant No.693683) and by the German Research Foundation (DFG) through the priority programme SPP 1726. We thank F. Smallenburg, M. Hermes, and L. Filion for useful discussions.

Author Contributions: Conceptualization, JdG; Methodology, CAV, CL, CB, and JdG; Software, CAV; Validation, CAV (lead) and CL and JdG (supporting); Investigation, CAV (lead) and CL (supporting); Original Draft, CAV and JdG; Review \& Editing, JdG, CB, CAV, and CL; Funding Acquisition, Resources, and Supervision, JdG and CB. The numerical code and analysis scripts used to obtain the data presented in this publication are available in Ref.~\cite{data}, along with a representative subset of the data.

\providecommand{\noopsort}[1]{}\providecommand{\singleletter}[1]{#1}%

\end{document}